\documentclass[conference]{IEEEtran}
\ifCLASSINFOpdf
  \usepackage[pdftex]{graphicx}
  \usepackage{cite}
\else
\fi
\usepackage{url}

\begin{document}
%
\title{



Universal End-to-End Neural Network for Lossy Image Compression

}


\author{\IEEEauthorblockN{Bouzid Arezki}
\IEEEauthorblockA{L2TI, Universite Sorbonne Paris Nord \\
99 avenue Jean-Baptiste Clément, \\ 93430 Villetaneuse, France \\
Email: bouzid.arezki@univ-paris13.fr}
\and
\IEEEauthorblockN{Fangchen Feng}
\IEEEauthorblockA{L2TI, Universite Sorbonne Paris Nord \\
99 avenue Jean-Baptiste Clément, \\ 93430 Villetaneuse, France \\
Email: fangchen.feng@univ-paris13.fr}
\and
\IEEEauthorblockN{Anissa Mokraoui}
\IEEEauthorblockA{L2TI, Universite Sorbonne Paris Nord \\
99 avenue Jean-Baptiste Clément, \\ 93430 Villetaneuse, France \\
Email: anissa.mokraoui@univ-paris13.fr}}


%


\maketitle

\begin{abstract}



This paper presents variable bitrate lossy image compression using a VAE-based neural network. An adaptable image quality adjustment strategy is proposed. The key innovation involves adeptly adjusting the input scale exclusively during the inference process, resulting in an exceptionally efficient rate-distortion mechanism. Through extensive experimentation, across diverse VAE-based compression architectures (CNN, ViT) and training methodologies (MSE, SSIM), our approach exhibits remarkable universality. This success is attributed to the inherent generalization capacity of neural networks. Unlike methods that adjust model architecture or loss functions, our approach emphasizes simplicity, reducing computational complexity and memory requirements. The experiments not only highlight the effectiveness of our approach but also indicate its potential to drive advancements in variable-rate neural network lossy image compression methodologies.
\end{abstract}


%
\IEEEpeerreviewmaketitle

\section{Introduction}
Neural network image compression has emerged as a superior alternative to conventional methods such as JPEG2000, offering distinct advantages~\cite{balle2018variational,Learning_Convo,CondiProModel,NEURIPS2018_53edebc5,lee2018contextadaptive,9190935,SwinNPE}. Neural architectures based on VAE (Variational AutoEncoder) typically consists of two main components: an encoder that transforms input image into a latent distribution, and a decoder that generates reconstructed image from this latent distribution. These two components work together to enable the learning of a compact and general representation of the input image. To achieve the best performance, the VAE aims to minimize the distortion between the original image and its compressed-decompressed version while adhering to a specified target bitrate constraint. This complex rate-distortion optimization problem is solved through the Lagrange formalism, introducing a Lagrange multiplier into the objective loss function of the VAE. This function typically consists of two terms: a distortion term, measuring the difference between the original image and the compressed-decompressed image, and a rate regularization term, imposing the bitrate constraint. 
This function, during the training process performed end-to-end, encourages the VAE to optimally adjust the parameters of both the encoder and the decoder, ensuring optimal compression performance while preserving the quality of the reconstructed image. Minimizing the objective function for a given value of the Lagrange multiplier allows obtaining the points on the convex hull corresponding to all possible rate-distortion points.  Therefore, it is essential to train the VAE for each combination  rate-distortion pair, corresponding to each value of the Lagrange multiplier. This leads to an individualized compression model for each specific value of the Lagrange multiplier. However, this strategy is restrictive as it requires training the VAE neural architecture each time, which can be time-consuming.
%
%
To overcome this, several strategies have been proposed in the scientific literature to achieve variable rates using a single trained model. For instance, Yoojin et al. introduced the conditional autoencoder that conditions the model on the Lagrange multiplier across the decoder, encoder, and entropy model~\cite{Yoojin}. Theis et al. introduced a scale parameter to fine-tune a pre-trained autoencoder for various rates~\cite{scale_botl}. Moreover, the approach  proposed in~\cite{MAE} extends this idea by integrating a modulated autoencoder with a VAE. Guerin Jr et al. suggested modifying the loss function to introduce rate control in VAE, though requiring the training of multiple models with fixed bitrates~\cite{constarined_loss}. Similarly, other approaches such as replacing the loss term with rate estimation or employing gain units for rate adaptation have been explored~\cite{loss_func,Asy_gain}. Furthermore, a variable quantization method controlled by the quantization bin size is also proposed to manage the bitrate~\cite{Yoojin}.

In these works, solutions typically involved altering the model architecture, adding modules conditioned by rate parameters, or adjusting quantization methods, often requiring additional fine-tuning or modification of the loss function~\cite{Yoojin,scale_botl,MAE,constarined_loss,Asy_gain}. Nevertheless, these strategies typically amplify the computational complexity and memory requirements of the network. Moreover, modifications to the loss function might result in reduced compression performance compared to traditional codecs such as BPG~\cite{loss_func}. Moreover, incorporating scale parameters in the latent space could potentially introduce compatibility issues with the model~\cite{scale_botl}.

This paper introduces a smart way to adjust image quality flexibly using a VAE-based neural network for variable bitrate lossy image compression. 
This method revolves around harnessing the capabilities of a singularly trained VAE model. The key lies in deftly adjusting the input scale within the image space exclusively during the inference process, resulting in a remarkably efficient rate-distortion mechanism. 
Through extensive experiments, we show that this approach works universally across various VAE-based image compression setups, including those using CNN, ViT, trained with MSE or SSIM as distortion metric in RGB color space. This strategy stems from the generalization capacity of neural networks. Furthermore, we argue that this method could spark progress in variable-rate neural network image compression techniques.

\section{End-to-end neural network image compression}
This section outlines the main features of a VAE for lossy image compression.
\subsection{Notations and Basic Concepts on VAE Frameworks}
In the image compression context, a VAE can represent an image in a compact form within a latent space, allowing its reconstruction with minimal loss of information and maintaining high visual quality. A VAE consists of three main components described below.

\textbf{Encoder} -- As in VAE-based image compression frameworks proposed in ~\cite{balle2018variational,Learning_Convo,CondiProModel,NEURIPS2018_53edebc5,lee2018contextadaptive,9190935,Yoojin,MAE,scale_botl,constarined_loss,
loss_func,Asy_gain,Gaussian_mixture,SwinNPE}, 
the encoder maps an input image $x$, using non-linear transforms denoted as $g_a(x,\phi_a)$ into a latent space $y$ representing the image. Another hyper latent $z$ is obtained by passing $y$ through a non-linear transform, referred to as \textit{"hyper encoder"} $h_a(y,\phi_h)$, where $\phi_a,\phi_h $ are the parameters of the generic parameterized transforms $h_a$ respectively in the encoder side $f_\theta () = \{g_a(;\theta_a), h_a(;\theta_h)\}$.

\textbf{Latent space} -- The quantizer transforms the latent space $z$ into discrete values. For the quantizer, uniform noise is added to the continuous latent before passing it to the prior model $\hat{z}=Q(z)$. Subsequently, $\hat{z}$ is entropy-coded with a learned factorized prior, which then passes through $h_s(\hat{z},\theta_h)$ to obtain $\mu$ and $\sigma$, the parameters of a factorized Gaussian distribution $P(y|\hat{z},\theta_h)= \mathcal{N}(\mu,,diag(\sigma))$ to model $y$. The quantized latent $\hat{y}=Q(y-\mu)+\mu$ is finally entropy-coded.

\textbf{Decoder} -- The non-linear transforms, denoted as \textit{"hyper decoder"} $h_s(\hat{z},\phi_h)$, receives $\hat{z}$ to deduce $\mu$ and $\sigma$ to model $y$. The non-linear transforms, denoted as \textit{"decoder"} $g_s(\hat{y},\phi_s)$, takes $\hat{y}$ to reconstruct the image $\hat{x}$, where $\phi_h, \phi_s$ are the parameters of the generic parameterized transforms $h_s$ and $g_s$ respectively in the decoder side $g_\phi () = {h_s(;\phi_h), g_s(;\phi_s)}$. 

Figure \ref{fig:pipeline} illustrates a generic end-to-end lossy image compression framework (see e.g. SwinNPE \cite{SwinNPE}).


\subsection{Rate-Distortion Optimization Formalism}
\begin{figure}[h]
\centering
 \includegraphics[width=0.35\textwidth]{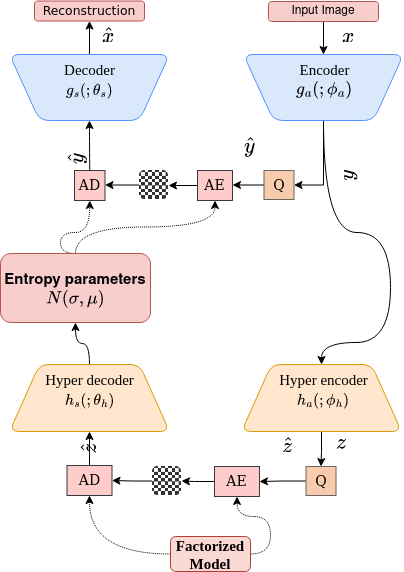}
  \caption{Generic architecture for image compression framework (e.g. SwinNPE \cite{SwinNPE}): $g_a$ and $h_a$, two modules in the right, present the encoder side of the VAE $f_\theta()$. $h_s$ and $g_s$, two modules in the left,  present the decoder side of the VAE $g_\phi()$, $y$ and $z$ are the latent vector, $Q$  the quantization module and $AE/AD$ present arithmetic encoding/decoding. }
 \label{fig:pipeline}
\end{figure}

The fundamental idea behind the rate-distortion optimization problem (RD) is to find an optimal configuration that minimizes the distortion subject to a constraint on the rate (or conversely, minimizes the rate subject to a constraint on the distortion). This tradeoff is often formalized using Lagrange multipliers and is expressed as an objective loss function that combines both rate and distortion terms.  To achieve this goal, the entire VAE image compression framework is trained end-to-end by seeking to minimize the following objective loss function: 
\begin{equation}
{L(\theta,\phi,\lambda)}= D (x, \hat{x}) + \lambda 
R(\hat{y}),   
\end{equation}
with $\lambda$ the Lagrange multiplier, $R$  the estimated bitrate, and $D$ the distortion between the original image $x$ and its compressed-decompressed version $\hat{x}$ given by $\hat{x} =   g_\phi (Q (f_\theta(x)))$ where $Q$ is the quantization operator. $\hat{y} = Q (f_\theta(x)) $ where $f_\theta ()$ concerns the encoder side $\{g_a(;\theta_a),  h_a(;\theta_h)\}$ and $g_\phi ()$ being the decoder side $\{g_s(;\phi_s), h_s(;\phi_h)\}$ (see Figure \ref{fig:pipeline}).

\section{Universal End-to-End VAE for Lossy image compression}
\label{proposed}
Remember that the VAE used for image compression learns to efficiently represent input data in a latent space by minimizing the objective loss function, which includes two main components: faithful reconstruction of input data (to reduce distortion) and regularization of the latent space (to control the rate). Regularization aims to prevent the latent space from becoming too complex, thus promoting more understandable and useful representations. The fact that all VAE parameters are computed in such a way to minimize the objective function suggests a holistic approach to model training. In other words, the VAE is specifically designed to optimize the rate-distortion pair, with a fixed regularization point. In summary, the model is tailored to find an optimal tradeoff between data compression and reconstruction quality, based on specific requirements set by the regularization point. To simplify the training process, this paper suggests using a single end-to-end trained VAE model for image compression, with a focus on minimizing the objective function.

Assume that the VAE, for a selected regularization point $\lambda_K$ for which the VAE achieves an excellent quality and a high bitrate (e.g. the point furthest to the right on the rate-distortion curve), has been trained to achieve an optimal rate-distortion ($D_K$, $R_K$) pair according to the minimization of the objective loss function. For this parametrization, the objective loss function is equal to: 
\begin{equation}
{L(\theta_K,\phi_K,\lambda_K)}= D_K + \lambda_K 
R_K, 
\end{equation}
where $ D_K = D (x, \hat{x})= D (x, g_{\phi_K} (Q (f_{\theta_K}(x))))$ and $ R_K = R(\hat{y}) = R(Q(f_{\theta_K}(x)))$.

We introduce a scaling factor, denoted $s$, belonging to the interval $]0,1[$. The original image $x$, intended for   compression, is then scaled by $s$ (i.e., $x_s=s \times x$) before being fed into the VAE for compression. This operation increases the distortion $D_K$. Indeed, along with the error induced by the already trained VAE (i.e., $D_K$), an additional error arises from scaling and rounding operations, as the compressed-decompressed image is deduced from  $\hat{x}' =  \lfloor \frac{ \hat{x}_s }{s}  \rfloor  $ ($\lfloor.\rfloor$ is the rounding operation). We can thus deduce that: 
$ D(x,\lfloor \frac{ \hat{x}_s }{s} \rfloor) > D_K=D(x,\hat{x})$. To preserve the objective loss function $ L(\theta_K,\phi_K,\lambda_K) $, a reduction in the bitrate is consequently enforced. Therefore a new rate-distortion pair ($D_{K_{s}}$, $R_{K_{s}}$) is inferred without the need to retrain the VAE compression architecture for an additional regularization point. 
By exploring all possible values of s ($]0,1[$), we succeed in constructing all the pair points on the rate-distortion curve using only a single VAE model. This will be discussed in the next section. 

\section{Experiments and Discussions}
This section shows that the strategy proposed in the previous section enables the construction of the complete rate-distortion curve with only one selected pair of points ($D_K, R_K$), using just a single  VAE model already trained for the corresponding regularization point $\lambda_K$. 

\subsection{Setup}
Different end-to-end neural network architectures have been selected: (i) Fully convolutional based methods: Factorized Hyperprior, Scale Hyperprior \cite{balle2018variational} and Joint Autoregressive and Hierarchical Priors \cite{NEURIPS2018_53edebc5}; (ii) Convolutional methods with attention modules: Discretized Gaussian Mixture Likelihoods and Attention Modules \cite{Gaussian_mixture}; and (iii) Transformer-based methods: SwinNPE \cite{SwinNPE}. 

The trained models are those provided by the CompressAI framework \cite{CompressAI}, including 8 different regularization values (i.e. $\lambda_i$), except for \cite{Gaussian_mixture} with 6 regularization values. All models are optimized based on Mean Squared Error (MSE) and Multi-Scale Structural SIMilarity (MS-SSIM) as distortion metric in RGB color space, and the quantization is performed using \textit{torch.round()}. 

In the case of SwinNPE \cite{SwinNPE}, we  trained the model on the CLIC2020 \cite{clic} dataset using 4 regularization values  $\lambda_1=0.003,\lambda_2=0.001,\lambda_3=0.0003,\lambda_4=0.0001$ and employed \textit{tf.round()} for quantization. The evaluation was performed on the Kodak dataset \cite{Kodak}. For \cite{Asy_gain}, we extracted the results from their paper.

\subsection{Discussions}
During the experimental process, we  select the regularization point $\lambda_K$ for which the VAE image compression has been trained to achieve an optimal rate-distortion ($D_K, R_K$) pair according to the minimization of the objective loss function. We define 9 scaling factors $s = 0.1,0.2...0.8,0.9$ belong to the interval $]0,1[$. 

\begin{figure}[h]
\centering
\includegraphics[width=0.45\textwidth]{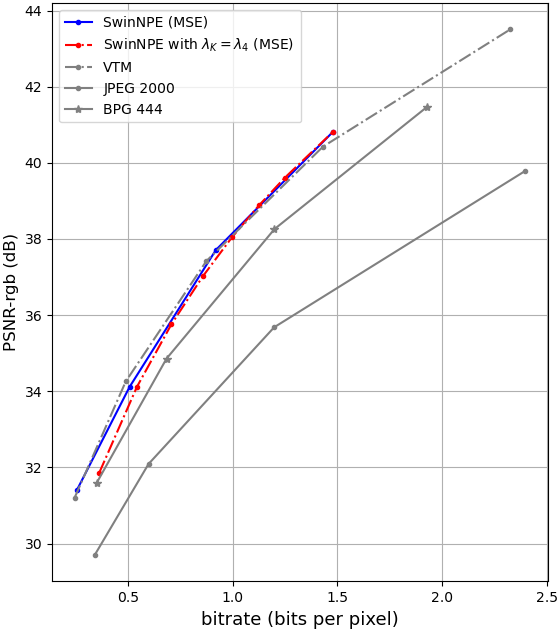}
\includegraphics[width=0.45\textwidth]{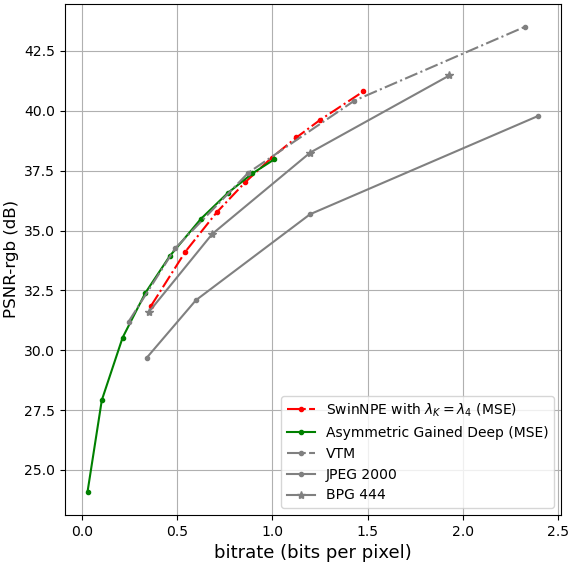}
\caption{1st Figure: Rate-distortion reference curve of SwinNPE \cite{SwinNPE} and dotted curve depicts the rate-distortion achieved by employing a single regularization point (i.e., the top-right point in the reference curve corresponding to a single trained SwinNPE model) with different values of the scaling factor $s$. 2nd Figure: the same dotted curve (i.e. using the top right point of SwinNPE) with Asymmetric Gained Deep variable bitrate model \cite{Asy_gain} on Kodak dataset \cite{Kodak}.}
\label{fig_curve}
\end{figure}

\begin{figure*}[h]
\centering
\includegraphics[width=0.9\textwidth]{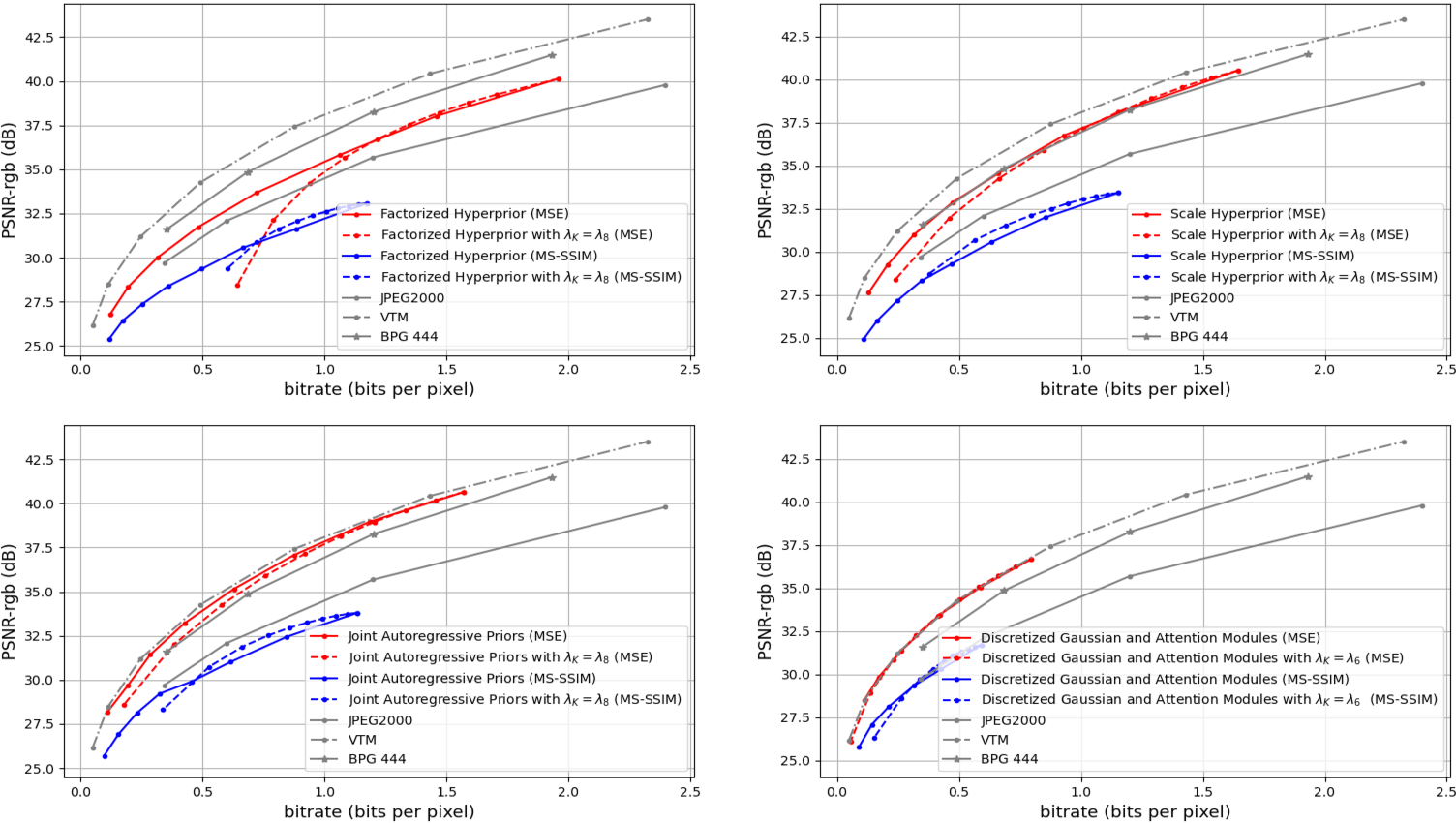}
\caption{Rate-distortion reference curves of Factorized Hyperprior, Scale Hyperprior\cite{balle2018variational}, Joint Autoregressive and Hierarchical Priors\cite{NEURIPS2018_53edebc5} and Discretized Gaussian Mixture Likelihoods and Attention Modules \cite{Gaussian_mixture}. 
The rate-distortion dotted curves are obtained by using a single regularization point (i.e., the top-right point in the reference curve corresponding to a single trained VAE model) with different values of the scaling factor $s$ on the Kodak dataset.
 \cite{kodak}.}
\label{sub_fig}
\end{figure*}

On the various plots, given in Figures \ref{fig_curve} and \ref{sub_fig}, the solid rate-distortion curves correspond to reference curves obtained when the model is trained with different regularization points and subsequently inferred on each trained model. The dotted rate-distortion curves correspond to curves obtained with a single trained model. For each new rate-distortion point, this trained model is inferred with an input image that has been scaled by a factor $s$, as explained in Section \ref{proposed}.

\begin{figure}[h]
\centering
\includegraphics[width=0.5\textwidth]{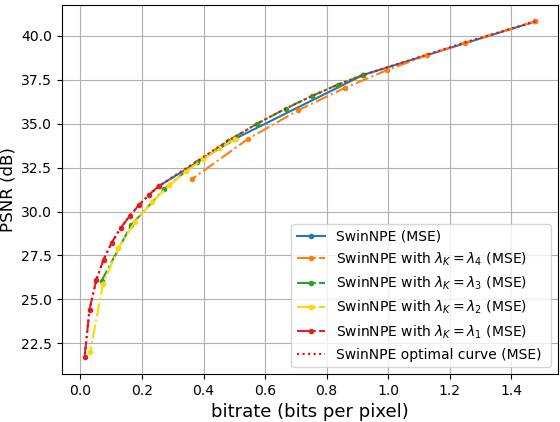}
\caption{The dotted curve, SwinNPE optimal curve, represents the envelope of the four rate-distortion curves obtained when successively exploiting the four SwinNPE trained models \cite{SwinNPE} (i.e. $\lambda_K = \lambda_1$, $\lambda_K = \lambda_2$, $\lambda_K = \lambda_3$, $\lambda_K = \lambda_4$) with different values of the scaling factor $s$ on the Kodak dataset \cite{kodak}.}
\label{expand}
\end{figure}

Figure \ref{fig_curve} depicts the rate-distortion reference curve of SwinNPE \cite{SwinNPE}. We choose $\lambda_K = \lambda_4$ and employ this trained SwinNPE to obtain the dotted rate-distortion curve, as explained below. This curve fits perfectly with the reference curve for a bitrate greater than $0.4$ bpp. 
A performance comparison  with the Asymmetric Gained Deep Image Compression With Continuous Rate Adaptation \cite{Asy_gain} shows that our strategy allows similar results. Figure \ref{expand} emphasizes the significance of selecting the trained SwinNPE model to act as a universal model. Indeed, it is observed that depending on the chosen regularization point (i.e., $\lambda_K=\lambda_1$, $\lambda_K=\lambda_2$, $\lambda_K=\lambda_3$, or $\lambda_K=\lambda_4$) the rate-distortion curve (called SwinNPE optimal curve in this figure) fits better to the reference curve while expanding the range of possible rates. Since the scale factor strategy increases distortion at the expense of reduced bitrate, relying on a model trained with a specific regularization point is crucial to achieving excellent quality of the   compressed-decompressed image while maintaining a high bitrate.

In Figure \ref{sub_fig}, the dotted distortion-rate curve was constructed based on the top-right regularization point, i.e., $\lambda_K$, of the corresponding solid reference curve for each image compression method. One can observe that our strategy allows for fitting the rate-distortion of the reference curves for Scale Hyperprior \cite{balle2018variational}, Joint Autoregressive and Hierarchical Priors \cite{NEURIPS2018_53edebc5}, and Discretized Gaussian Mixture Likelihoods and Attention Modules \cite{Gaussian_mixture}. However, for Factorized Hyperprior \cite{balle2018variational}, the performance decreases compared to the reference (up to $\approx1 dB$) when significantly reducing the dynamic range of the image (i.e., when $s$ is too small).

\begin{figure*}[h]
\centering
\includegraphics[width=0.9\textwidth]{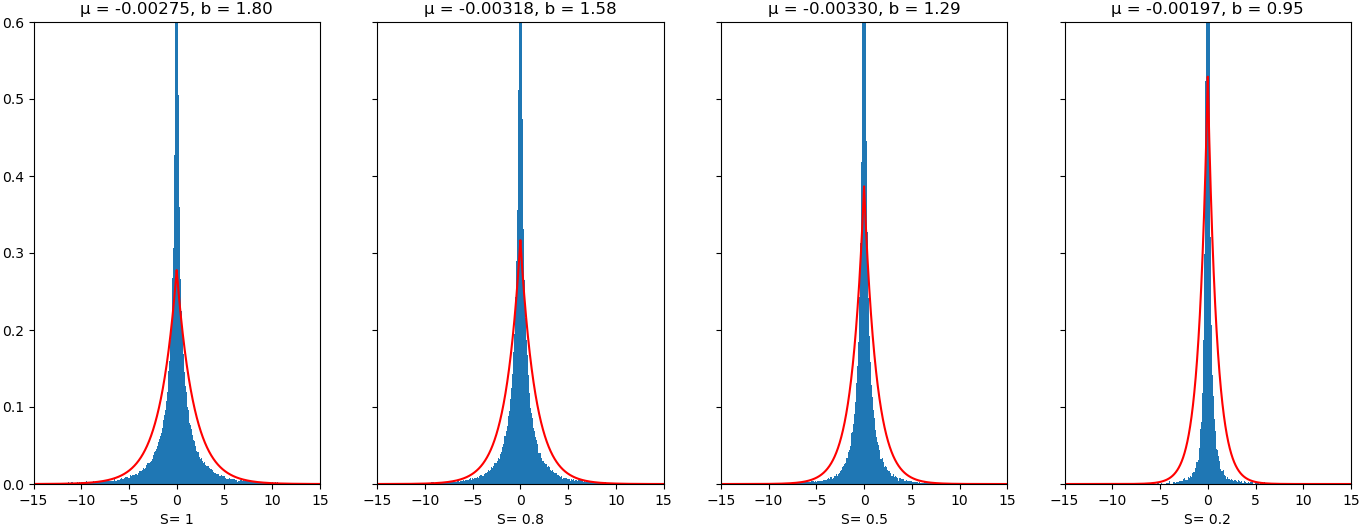}
\caption{Normalized histograms of the latent space of \textit{"kodim01.png"} image in Kodak dataset \cite{Kodak} with SwinNPE \cite{SwinNPE} using three scaling factors $s=0.2$, $s=0.5$, and $s=0.8$. }
\label{sub_fig_histo}
\end{figure*}

To grasp the impact of the scaling factor $s$, Figure \ref{sub_fig_histo} displays the normalized histograms of the latent space for SwinNPE \cite{SwinNPE}, using the \textit{kodim01.png} image from the Kodak dataset \cite{Kodak}. The histograms adhere to a Laplacian distribution that contracts as the scale $s$ decreases, with an amplified amplitude indicating an increased sparsity level in the latent space. This behavior confirms the explanation provided in Section \ref{proposed}.

The common feature of the VAE used in this paper is that quantization is fixed, and the parameters of this quantization are not fine-tuned during the learning process. This imparts a level of flexibility to the regularization point, influencing the tradeoff between rate and distortion. Introducing the scale factor is akin to incorporating an additional form of uniform quantization, which increases the distortion according to the rounding step to recover the original image. It would be interesting to explore other scale factor variations, such as a non-uniform scale factor, in future research.

\section{Conclusion}
This paper proposes variable bitrate image compression through a VAE-based neural network.  Our innovative strategy, based on a uniquely trained VAE model, skillfully fine-tunes the input image scale during the inferred process, resulting in an efficient rate-distortion mechanism.  Our approach is applicable to various VAE-based image compression frameworks, including those using CNNs and attention mechanisms. We argue that this has the potential to advance techniques for variable-rate neural network image compression by reducing complexity and memory requirements.

Future research will  evaluate our technique using perceptual metrics, comparing it to various perceptual VAE-based methods across different datasets.




%

\end{document}